\documentclass[%
reprint,
superscriptaddress,
 showpacs,preprintnumbers,
 amsmath,amssymb,
 aps,
prl,
 floatfix,
]{revtex4-2}

\usepackage{booktabs}
\usepackage{array}
\newcolumntype{L}[1]{>{\raggedleft\arraybackslash}p{#1}}
\AtBeginDocument{
\heavyrulewidth=.08em
\lightrulewidth=.05em
\cmidrulewidth=.03em
\belowrulesep=.65ex
\belowbottomsep=0pt
\aboverulesep=.4ex
\abovetopsep=0pt
\cmidrulesep=\doublerulesep
\cmidrulekern=.5em
\defaultaddspace=.5em
}

\usepackage{subcaption}
\usepackage{soul}
\usepackage{hyperref}
\usepackage{color}
\usepackage{xcolor}
\usepackage{graphicx}% Include figure files
\usepackage{dcolumn}% Align table columns on decimal point
\usepackage{bm}% bold math
\usepackage{cancel}
\usepackage[utf8]{inputenc}
\usepackage{upgreek}

\providecommand{\abs}[1]{\vert #1\vert}
\newcommand{\etal}{\textit{et al}. }
\newcommand{\ie}{\textit{i}.\textit{e}.}

\begin{document}

% \title{Using a multimode analysis to extract magnetism from single linear-polarization X-ray ptychography}
% Magnetic domain imaging using a single polarization and multimodal-ptychography
% multimodal magnetic Ptychography imaging using a single polarization
% Polarization-resolved X-ray imaging without a physical polarization analyzer
% Polarization-resolved X-ray imaging using multimodal-ptychography
% imaging the magnetization with single linear polarization ptychograpy by using a multimode analysis
\title{Magnetic X-ray imaging using a single polarization and multimodal-ptychography}

\author{Marisel Di Pietro Martínez}
\email{Marisel.DiPietro@cpfs.mpg.de}
\affiliation{Univ. Grenoble Alpes, CNRS, Grenoble INP, SIMaP, 38000 Grenoble, France}
\affiliation{Max Planck Institute for Chemical Physics of Solids, Noethnitzer Str. 40, 01187 Dresden, Germany}
\author{Alexis Wartelle}
\affiliation{Univ. Grenoble Alpes, CNRS, Grenoble INP, SIMaP, 38000 Grenoble, France}
\affiliation{European Synchrotron Radiation Facility, F-38043 Grenoble, France}
\affiliation{Université Grenoble Alpes, CNRS, Institut Néel, 38042 Grenoble, France}
\author{Nicolas Mille}
\affiliation{Synchrotron SOLEIL, Saint Aubin, BP 48, 91192 Gif-sur-Yvette, France}
\author{Stefan Stanescu}
\author{Rachid Belkhou}
\affiliation{Synchrotron SOLEIL, Saint Aubin, BP 48, 91192 Gif-sur-Yvette, France}
\author{Farid Fettar}
\affiliation{Université Grenoble Alpes, CNRS, Institut Néel, 38042 Grenoble, France}
\author{Vincent Favre-Nicolin}
\affiliation{European Synchrotron Radiation Facility, F-38043 Grenoble, France}
\author{Guillaume Beutier}
\affiliation{Univ. Grenoble Alpes, CNRS, Grenoble INP, SIMaP, 38000 Grenoble, France}

\date{\today}% It is always \today, today,
             % but any date may be explicitly specified

\begin{abstract}
%%% Abstract <600 characters
Polarized X-rays allow for imaging birefringent or dichroic properties of materials with nanometric resolution.
To disentangle these properties from the electronic density, either a polarization analyzer or several measurements with different polarizations (typically two, or more) are needed.
Here we demonstrate that ptychography can disentangle these from a single-polarization measurement
% % Using X-ray ptychography, w
% We imaged samples in which two X-ray-matter interactions with different polarization properties occur: the linear polarization of the incident X-rays is conserved by one of them, while it is rotated by 90$^\circ$ by the other one. 
% The former provides a chemical image of the sample, and the latter a magnetic image. Both exit waves, with mutually orthogonal polarizations, are retrieved without a physical analyzer 
by using a multimodal analysis.
% of the ptychographic data.
% We prove the method in a magnetic sample for which we obtain an image of the magnetic domains from a single-polarization ptychographic measurement.
% We validate the results by comparing them to data obtained using circular dichroism.
% \ma{Example of future application}
This new method provides an alternative
% the use of a physical analyzer, which would induce optical aberrations and increase tremendously the acquisition times, 
to obtain polarization-resolved images of a sample when manipulating the incident polarization is not possible nor sufficient.
\end{abstract}

\maketitle

%##################################################################################
\section{Introduction}
\label{sec:intro}

% Motivation, state of the art, what is lacking in terms of experimental methods, what people have done so far with circular polarization and what has not been done and what we propose. Advantages.

Since their discovery, X-rays have been recognised for their potential in microscopy~\cite{Röntgen}. 
Many imaging modes have been developed, in full-field or in scanning mode, with or without lenses, using various contrast mechanisms such as absorption \cite{Röntgen}, fluorescence \cite{Vogt}, phase contrast \cite{Endrizzi}, diffraction \cite{Chahine,Simons}, etc. 
In particular, using polarized X-rays and combining measurements with different polarizations allows for evidencing the birefringence \cite{Palmer,Gao} and dichroism \cite{Fischer,Platunov,Ade} of materials, whether they are of structural \cite{Platunov,Palmer,Gao} or magnetic \cite{Fischer,Stohr} origin. 
For instance, linear birefringence or dichroism is obtained by combining measurements with two orthogonal linear polarizations, while circular birefringence or dichroism is obtained by combining measurements with two orthogonal circular polarizations.

While controlling the polarization of the incident photons is nowadays fairly easy, in particular with synchrotron radiation \cite{Sasaki,Giles}, analysing the polarization of X-rays after their interaction with the sample remains a difficult task. This is particularly true for most imaging methods, since they rely on the use of 2D detectors (with the notable exception of scanning microscopies that can be performed with a point detector). The usual way to analyze the polarization of X-rays is to diffract them at 90$^\circ$ from a crystal or an artificial grating with a period tuned for their wavelength, according to Bragg's law \cite{Chandrasekhar,Kortright}: an azimuthal scan of the analyzer around the propagation axis allows to fully determine the Stoke polarization parameters \cite{Wang2012pol}. This procedure assumes a well defined propagation axis, which is not always the case when measuring with an area detector pointing at a small sample. When the angular spread of X-rays exceed the angular acceptance of the analyzer, its efficiency becomes inhomogeneous, altering the quality of recorded images.
It is nevertheless of great interest to perform a polarization analysis of X-rays after the sample since the change of polarization in the sample carry important information~\cite{Mazzoli,Wang}.
% \textbf{Specific examples?}

Ptychography is among the most recent imaging methods \cite{hoppe1970principles,Faulkner,rodenburg2007hard,thibault2008high,PhysicsToday,donnelly2016high}. It is based on the combination of direct-space scanning and reciprocal-space measurements with a coherent beam. It benefits from high-resolution and retrieves 
the complex exit wave allowing for both amplitude and phase contrast.
% amplitude as well as phase contrast. 
Interestingly, thanks to the redundancy of information measured in a ptychographic data set, more information than a single image of the sample can be recovered. This property is often also used to retrieve an image of the probe illuminating the sample~\cite{thibault2009probe}, or the true positions of the beam with respect to the sample during the scan \cite{maiden}.

Going further, Thibault \etal pointed out that ptychography is able to disentangle several waves exiting the sample without interfering, \ie{} mutually incoherent, which reduce the interference contrast in the scattering patterns measured at each point of the scan \cite{thibault2013reconstructing}. This incoherence can be either due to the partial coherence of the incident beam, a very common case, but also to a mixture of orthogonal states in the sample. The key idea is that X-rays exiting the sample can be considered as a mixture of coherent but mutually incoherent waves, regardless of the reason why they are not interfering. It is for instance the case when the sample fluctuates at timescales below the timescale of the measurements~\cite{Clark}. Another possible reason for two electromagnetic waves not to interfere is if they have orthogonal polarizations. The latter can be linear, or circular, or more generally, elliptical. Therefore, ptychography should be able to disentangle exit waves with orthogonal polarizations, overcoming the aforementioned usual limitations in the X-ray polarization analysis.
To demonstrate this, here we show the case of magnetic materials in which the magnetic texture can be disentangled from the electronic structure,
% chemical structure
from a single ptychographic measurement with incident linear polarization.

%The purpose of this paper is to demonstrate the capability of ptychography to resolve orthogonal polarizations with a single measurement. If exit waves with orthogonal polarizations can be imaged while the sample is illuminated by X-rays with a well-defined polarization, then important information on the polarization properties of the sample can be extracted. Specifically, we target magnetic materials and we aim at retrieving two disentangled images, one for the magnetization and another one for the electronic structure, from a single ptychographic measurement.

\section{Multimodal formulation}

% \gb{I'VE REMOVED SOME BLABLA HERE, WHICH IS REDONDANT WITH THE INTRODUCTION.}
%Ptychography is a diffraction-based microscopy technique which consists on scanning an extended sample with a shifting illumination to get redundant reciprocal-space information to reconstruct a highly-resolved real-space image~\cite{hoppe1970principles,hoppe1982trace}. 
% coherent scattering
%\gb{I WOULD REMOVE THE FOLLOWING SENTENCE:}{since resolution is not the point of the paper}
%In particular, the use of coherent X-rays allows the achievement of nanometric resolution, down to $10$\,nm~\cite{rodenburg2007hard,thibault2008high,PhysicsToday}.

\begin{figure}[!b]
\begin{center}
  \resizebox{\columnwidth}{!}{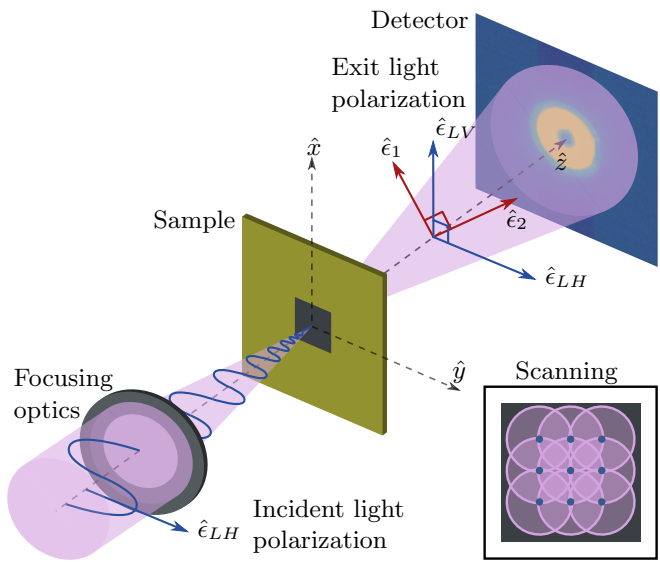}
\end{center} 
\caption{Ptychography setup: Coherent X-rays illuminate an area of the sample and a diffraction pattern is measured in the far field by a detector. The sample is scanned following a mesh that ensures overlapping illumination, as illustrated in the lower right inset, providing the data redundancy required for a robust image reconstruction.
Polarized light provides contrast for dichroic or birefringent samples.
$\hat{\epsilon}_{LH}$ and $\hat{\epsilon}_{LV}$ represent the unity vectors for linearly horizontal and vertical polarized light, respectively, while $\hat{\epsilon}_1$ and $\hat{\epsilon}_2$ represent an alternative  orthogonal base for the polarizations.}
\label{fig:ptycho} 
\end{figure}

% via an iterative retrieval algorithm that deals with the phase problem.
% To obtain the amplitude and phase of the object in real space, iterative algorithms are used.
% advantages of a wide field of view and fast convergence, quantitative images
% ‘probe’ beam,  recording a set of diffraction patterns, each corresponding to a different specimen position.
% ptychography’s high resolution images are complex, mapping both
% the amplitude and phase of the specimen transmission function.
In ptychography, for each positioning of the beam relative to the sample, a diffraction pattern is measured with the detector, as illustrated in Fig.~\ref{fig:ptycho}.
The key of this technique is the overlapping of the illuminated areas (see Fig.~\ref{fig:ptycho}, right-lower corner). This results in redundancy of the data which allows a robust reconstruction of a wide field of view of the sample.
The resulting set of diffraction intensities can be used to recover the complex transmitted or exit wave.

To derive the multimodal formulation in this context, let us calculate the transmission of a polarized incident wave $\vec{E}_0 = \Psi_0 \hat\epsilon$ defined by its complex amplitude $\Psi_0$ and its polarization $\hat\epsilon$. When $\hat\epsilon$ is an eigenvector of the polarization-dependent transmission function, \ie{} the polarization is not modified through the sample, the transmitted complex amplitude can be written as: %$\vec{E} = (\Psi_1,\Psi_2)$ defined by its scalar amplitudes $\Psi_1$ and $\Psi_2$ on an arbitrary orthogonal polarization basis $(\hat\epsilon_1,\hat\epsilon_2)$. In general, the amplitude of such wave can be written as
\begin{equation}
    \Psi = \Psi_0 \exp{\left( \frac{2\pi i}{\lambda} \int n \mbox{d}z\right)},
    \label{eq:exitwave}
\end{equation}
where %$\Psi_0$ is the incident wave, 
$\lambda$ is the wavelength, 
% $\hat\epsilon$ is the polarization of the light,
the wave propagates along the $z$ axis,
and $n=n(x,y,z)$ is the optical index. 
% amplitude and phase of the transmission function.
When $\hat\epsilon$ is not an eigenvector of the polarization-dependent transmission function, the wave can always be decomposed on a polarization basis of eigenvectors $\vec{E} = \Psi_1 \hat\epsilon_1 + \Psi_2 \hat\epsilon_2$ such that Eq.~\ref{eq:exitwave} applies independently to both components with different optical indices.

In ferromagnetic materials with isotropic local structure, circular polarizations are such eigenvectors.
% When the incident wave is circularly polarized, 
The optical index can then be expressed as $n = n_c \pm n_m$, where $n_c$ and $n_m$ denote the contribution from the electronic and the magnetic structure, respectively, and the change in sign depends on the helicity of the polarization. 
Therefore, in this case, Eq.~\ref{eq:exitwave} leads to %we can write 
\begin{equation}
%    \Psi^{\pm} = \Psi_0 \exp{ \left(\int \frac{2\pi i n_c z}{\lambda} \mbox{d}z \right)}  \exp{\left( \int \pm \frac{2\pi i n_m z}{\lambda} \mbox{d}z \right)}, \label{eq:pm}
    \Psi^{\pm} = \Psi_c \exp{\left( \pm \frac{2\pi i}{\lambda} \int n_m \mbox{d}z \right)}, \label{eq:pm}
\end{equation}
where $\Psi^{+}$ and $\Psi^{-}$ correspond to the exit waves for right and left circular polarization, respectively, and 
\begin{equation}
    \Psi_c = \Psi_0 \exp{\left( \frac{2\pi i}{\lambda} \int n_c \mbox{d}z\right)}.
\end{equation}
% \begin{equation}
% \Psi_1 = \Psi_0 \exp{ \int \frac{2\pi i n_c z}{\lambda} \mbox{d}z}.
% \end{equation}

We can isolate the magnetic part of the optical index by calculating the ratio 
$\frac{\Psi^+}{\Psi^-}$, which leads to:
%\begin{equation}
%    \frac{\Psi^+}{\Psi^-} = \exp{\int  \frac{4\pi i n_m z}{\lambda} \mbox{d}z}.
%\end{equation}
%Since the optical index is a complex quantity, $n = 1-\delta-i\beta$, we can access the real and imaginary parts by calculating
\begin{eqnarray}
    \mbox{arg} \left(\frac{\Psi^+}{\Psi^-}\right) \propto \delta_m, \label{eq:phase}\\
    \ln \left|\frac{\Psi^+}{\Psi^-}\right| \propto \beta_m. \label{eq:amp}
\end{eqnarray}
where $\delta_m$ and $\beta_m$ are the real and imaginary parts of $n_m$, respectively.
% Therefore, a magnetic image can be obtained from the absorption contrast (Eq.~\ref{eq:amp}) or the phase contrast (Eq.~\ref{eq:phase}).
Eq.~\ref{eq:phase} corresponds to what is known as the phase contrast and Eq.~\ref{eq:amp} corresponds to the amplitude contrast.
Similarly, we can access the electronic contribution to the optical index
by calculating the product ${\Psi^+}{\Psi^-}$, which results in
\begin{eqnarray}
    \mbox{arg} \left(\Psi^+  \Psi^-\right) \propto \delta_c + \mbox{cte}, \label{eq:phase_c}\\
    \ln \left|\Psi^+\Psi^-\right| \propto \beta_c  + \mbox{cte},
    \label{eq:amp_c}
\end{eqnarray}
where $\delta_c$ and $\beta_c$ are the real and imaginary parts of $n_c$, respectively.

Combining ptychography in the small angle scattering geometry with X-ray magnetic circular dichroism (XMCD) allows for probing either $\delta_m$ or $\beta_m$ which in turn are proportional to the magnetic component that is parallel to the beam direction, that is $m_z$ according to the reference frame defined in Fig.~\ref{fig:ptycho}.
XMCD-ptychography consists on performing one measurement for each helicity of the circular polarization~\cite{Shi2016,donnelly2016high,donnelly2017three}, 
and then, by using Eqs.~\ref{eq:phase} or \ref{eq:amp}, an image of the magnetization can be obtained.

In general, we can decompose any polarization vector onto an orthogonal basis of circular polarizations.
In particular, in the case of linearly polarized light, this means that we can
write the exit wave as the sum of two circularly polarized plane waves
% \begin{eqnarray}
%     \Psi_0 \hat\epsilon_{L1} = \frac{1}{2} \Psi_0 \left(\hat\epsilon_{L1} + i \hat\epsilon_{L2} \right) + \frac{1}{2} \Psi_0 \left(\hat\epsilon_{L1} - i \hat\epsilon_{L2} \right),
% \end{eqnarray}
\begin{eqnarray}
    \vec{E} = \frac{\sqrt{2}}{2} \left[ \Psi^+ \left(\frac{\hat\epsilon_{L1} - i \hat\epsilon_{L2}}{\sqrt{2}}\right) + \Psi^- \left(\frac{\hat\epsilon_{L1} + i \hat\epsilon_{L2}}{\sqrt{2}}\right)\right],
\end{eqnarray}
where $\hat\epsilon_{L1}$ and $\hat\epsilon_{L2}$ are two orthogonal linear polarizations and the combination $\left(\hat\epsilon_{L1} - i \hat\epsilon_{L2} \right)/\sqrt{2}$ and $\left(\hat\epsilon_{L1} + i \hat\epsilon_{L2} \right)/\sqrt{2}$ gives us the eigenvectors for right and left circular polarizations, respectively.
% Considering Eq.~\ref{eq:pm}, w
We can rewrite the previous expression as
% \begin{widetext}
% \begin{equation}
% % \begin{eqnarray}
% % \begin{split}
%     \vec{E} = \frac{1}{2} \Psi_c \left[ \left( \exp{ \int \frac{2\pi i n_m z}{\lambda} \mbox{d}z} + \exp{ \int \frac{- 2\pi i n_m z}{\lambda} \mbox{d}z} \right) \hat\epsilon_{L1} \right. -
%     \left. i \left( \exp{ \int \frac{2\pi i n_m z}{\lambda} \mbox{d}z} - \exp{ \int \frac{- 2\pi i n_m z}{\lambda} \mbox{d}z} \right) \hat\epsilon_{L2} \right].
%     \label{eq:lin}
% % \end{split}
% % \end{eqnarray}
% \end{equation}
% \end{widetext}
% \begin{widetext}
\begin{equation}
% \begin{eqnarray}
% \begin{split}
    \vec{E} = \Psi_c \left[ \left(\frac{\Psi^+ + \Psi^-}{2}\right) \hat\epsilon_{L1} 
    + \left(\frac{\Psi^+ - \Psi^-}{2i}\right) \hat\epsilon_{L2} \right].
    \label{eq:lin}
% \end{split}
% \end{eqnarray}
\end{equation}
% \end{widetext}
%where
%\begin{equation}
%    \Psi_c = \Psi_0 \exp{ \int \frac{2\pi i n_c z}{\lambda} \mbox{d}z}.
%\end{equation}

% Given a sample of thickness $d$,
% % in the first Born approximation (
% if the sample is magnetically optically thin, that is $n_m d \ll \lambda$, we can approximate Eq.~\ref{eq:lin} to
% \begin{equation}
%     \vec{E} \approx \Psi_c \left(\hat\epsilon_{L1} + \int \frac{2\pi n_m z}{\lambda} \mbox{d}z \,\hat\epsilon_{L2}\right).
% \end{equation}
% Two conclusions emerge from this result. Firstly, 
Since the linear polarizations $\hat\epsilon_{L1}$ and $\hat\epsilon_{L2}$ are orthogonal, their amplitudes
% \begin{equation}
%     \Psi^{L1} = \Psi_c \mbox{ and }
%     \Psi^{L2} = \Psi_c \int \frac{2\pi n_m z}{\lambda} \mbox{d}z, \label{eq:2modes}
% \end{equation}
\begin{eqnarray}
    \Psi^{L1} &=& \Psi_c \cos{\left( \frac{2\pi}{\lambda} \int n_m \mbox{d}z\right)} \label{eq:2modes_1} \mbox{ and }\\
    \Psi^{L2} &=& \Psi_c \sin{\left( \frac{2\pi}{\lambda} \int n_m \mbox{d}z\right)} \label{eq:2modes_2}
\end{eqnarray}
do not interfere.
Therefore the intensity measured in an experiment with linear polarization can be expressed as the sum of the intensities of these two independent modes\begin{equation}
 I = \abs{\mathcal{F}\{\Psi^{L1}\}}^2 + \abs{\mathcal{F}\{\Psi^{L2}\}}^2,
 \label{eq:I2modes}
\end{equation}
where $\mathcal{F}$ denotes the Fourier transform operator.
This result has been previously used in Ref.~\cite{tripathi2011dichroic}, for example, to isolate the magnetic term by subtracting two measured diffraction intensities: one with the sample magnetically saturated and another one with the sample not saturated.
However, the fact that the two contributions to the intensity can be summed directly suggests they could also be 
% treated as two orthogonal modes of the object in the context of a ptychographic analysis.
% which means they can be 
recovered with a multimodal analysis of the ptychography data.

% For linearly polarized light, the scattering factor in the the small angle regime results in
% \begin{equation}
%  f = f_c \left( {\begin{array}{cc}
%     1 & 0 \\
%     0 & 1 \\
%   \end{array} } \right) - i f_m^{(1)} m_z \left( {\begin{array}{cc}
%     0 & 1 \\
%     1 & 0 \\
%   \end{array} } \right),
% \end{equation}
% % \textbf{higher orders many times weaker: mxmy}
% where we have considered the base for linear polarization $(\hat{\epsilon}_{LH},\hat{\epsilon}_{LV})=\left(\begin{pmatrix} 1 \\ 0 \end{pmatrix},\begin{pmatrix} 0 \\ 1 \end{pmatrix}\right)$.
% % and higher orders are typically negligible.
% Notice that in this case the electronic density and the magnetism are orthogonal. This means that the intensity can be expressed as
% \begin{equation}
%     % I(\textbf{q}) = \abs{\mathcal{F}\{\Psi_k(f_c)\}}^2 + \abs{\mathcal{F}\{\Psi_k(f_m^{(1)} m_z)\}}^2.
%  I(\textbf{q}) = \abs{\mathcal{F}\{f_c(\textbf{r})\}}^2 + \abs{\mathcal{F}\{f_m^{(1)}(\textbf{r}) m_z\}}^2.
%  \label{eq:I2modes}
% \end{equation}

% To this end, we consider a multimode analysis. 
% In this context, the probe and the object in Eq.~\ref{eq:psi} can be written as vectors, $\vec{P}=(P_1, ..., P_M)$ and $\vec{O}=(O_1,...,O_N)$, with each element representing a mode. 
The multimodal approach has been previously shown to be useful
% of the probe allows 
for disentangling mixed states of the beam for instance~\cite{thibault2013reconstructing}, 
% when considering a probe with several modes
% while multiple object modes have also been previously used 
as well as to reconstruct non-orthogonal states in a birefringent liquid crystal with the aid of prior knowledge about the sample~\cite{li2016breaking}.
Here we use the multimodal analysis to exploit the orthogonality of the two modes from linear polarization, given by Eqs.~\ref{eq:2modes_1} and \ref{eq:2modes_2},
% of the magnetic and the electronic structure with linear polarization,
to resolve both the magnetic and the electronic structure from a single ptychographic measurement without any extra constrain.

% While the first mode $\Psi^{L1}$ gives us directly the contrast for $n_c$,
Calculating the ratio between the modes from Eqs.~\ref{eq:2modes_1} and \ref{eq:2modes_2},
% \begin{equation}
%     \frac{\Psi^{L2}}{\Psi^{L1}} = - \int \frac{4\pi n_m z}{\lambda} \mbox{d}z,
%     % \label{eq:2modes}
% \end{equation}
and comparing this result to Eqs.~\ref{eq:phase} and ~\ref{eq:amp}, we find the relations
% \begin{eqnarray}
%     2 \mbox{Re}\left(\frac{\Psi^{L2}}{\Psi^{L1}}\right) &=& \mbox{arg} \left(\frac{\Psi^+}{\Psi^-}\right) \propto \delta_m, \label{eq:imag}\\
%     2 \mbox{Im}\left(\frac{\Psi^{L2}}{\Psi^{L1}}\right) &=& \ln \left|\frac{\Psi^+}{\Psi^-}\right| \propto \beta_m, \label{eq:real}
% \end{eqnarray}
\begin{eqnarray}
    2 \mbox{Re}\left[\tan^{-1}\left(\frac{\Psi^{L2}}{\Psi^{L1}}\right)\right] &=& \mbox{arg} \left(\frac{\Psi^+}{\Psi^-}\right) \propto \delta_m, \label{eq:imag}\\
    2 \mbox{Im}\left[\tan^{-1}\left(\frac{\Psi^{L2}}{\Psi^{L1}}\right)\right] &=& \ln \left|\frac{\Psi^+}{\Psi^-}\right| \propto \beta_m, \label{eq:real}
\end{eqnarray}
while computing the squared sum of the modes leads to
\begin{equation}
    \left(\Psi^{L1}\right)^2 + \left(\Psi^{L2}\right)^2 = \Psi^+  \Psi^- = \Psi_c^2.
\end{equation}
These equations allow to compare the modes obtained from the single linear polarization with the usual XMCD signal.

% Peng Li: here we show successful reconstruction of all of the states in an optical experiment that involves a dynamic object composed of two states (not mutually orthogonal) and an illuminating probe with five states.

\section{Experimental details}

We test the method experimentally on two different samples. We use an Fe/Gd multilayer with nominal stacking of Ta($7$)/[Fe($0.7$)/Gd($0.7$)]$_{60}$/Ta($7$) (thickness in nm),
% , leading to a nominal average composition of Fe$_{74}$Gd$_{26}$ and a total stack thickness of $84$\,nm.
% This sample was measured at the Fe $L_3$-edge at energy $708.36$\,eV.
% We also study 
and a much thinner sample consisting on a Co/Pt multilayer with nominal stacking of [Co($0.6$)/Pt($1$)]$_{10}$ (thickness in nm).
% This sample was measured at the Co $L_3$-edge at energy $782.5$\,eV.
Both the Fe/Gd and the Co/Pt multilayers were grown on a $300$\,nm-thick Si$_3$N$_4$ membrane suitable for transmission X-ray measurements,
and were measured at the peak of absorption at the Fe and Co $L_3$-edge, respectively.

The ptychography data presented in this work was acquired at HERMES beamline, at SOLEIL synchrotron.
Here polarized X-rays are delivered by an helicoidal undulator, allowing to easily switch from horizontal polarization to either of both circular polarizations, with a polarization rate close to $100$\%. The energy of the photons was tuned by a grating monochromator. The coherence of the beam was ensured by a set of apertures in front of the endstation. The coherent beam was focused using a Fresnel zone plate with $50$\,nm outer zone width. The sample was placed a few microns downstream the focus in order to obtain a structured illumination of ~$100$\,nm. The small angle coherent diffraction patterns are acquired on an adapted Tucsen Dhyana 95 sCMOS camera~\cite{desjardins2020backside} with $2048\times2048$ pixels of $11$\,$\upmu$m. Typical ptychographic datasets were obtained in grid scans with step size of $25$\,nm and acquisition times of $50$\,ms per point.
The geometrical settings were such that the pixel size of the direct space images is $7$\,nm.
% at the Co $L_3$ edge.

% \textbf{What quantities are we presenting.}

To analyze the ptychography data, we use PyNX, a Python-based high performance toolkit for coherent X-ray imaging data~\cite{favre2020pynx}.
We use a single probe and an object with two modes. The first mode is initialized with random positive real numbers and the second one with random complex numbers. The amplitude of the second mode is set initially to $1/10$ times the amplitude of the first mode to induce a better separation of the charge and magnetic part into the first and second modes, respectively, as it is suggested by simulations in Ref.~\cite{zhang2017quantitative} to disentangle mixed-states.
% 1. Load all the ptycho scans.
% 1. get the probe for 1 object.
% 2. Add second mode to object.
% We define an initial object with two modes, one 
% obj_init_info = {'type': 'random', 'range': (0.9, 1, 0, 0.5), 'shape': (nxy, nxy)}
% n = abs(ob1[0]).mean() / 10
%     ob1[i] = np.random.uniform(0, n, (nxy, nxy)) * np.exp(1j * np.random.uniform(0,2*np.pi, (nxy,nxy)))
% 1 probe 
% Then, we run $150$ iterations of the alternating projections algorithm.
Then, we run $100$ iterations of the difference map algorithm, $50$ iteration of the alternating projections algorithm, and finally $150$ iterations of the maximum likelihood algorithm.
% DM**100,AP**50,ML**150
The reconstruction is achieved by iterating between the real-space estimated object
% , in this case $O_1$ and $O_2$, 
and the Fourier-space where the amplitude is constrain to the measured diffraction intensity.

%  Energy, position, size um, scan points, time ms
% All same size, for CL, CR, LH
%086: SampleImage(fine, 782.5, 90, 5921.5, 1, 1, 40, 40 ,50, True)
%089: SampleImage(fine, 782.5, 90, 5916, 1, 1, 40, 40 ,50, True)
%096: SampleImage(fine, 782.5, 90, 5901, 1, 1, 40, 40 ,50, True)
% The CL, CR of 002-003 were 1.5x1.5um.
%005: SampleImage(fine, 708.36, -444.3, 5552.1, 0.8, 0.8, 40, 40 ,50, True)

\section{Experimental demonstration}
% \section{Reconstruction case with uniform charge structure}
\label{sec:results}

\begin{figure}[!b]
	\includegraphics[width=\columnwidth]{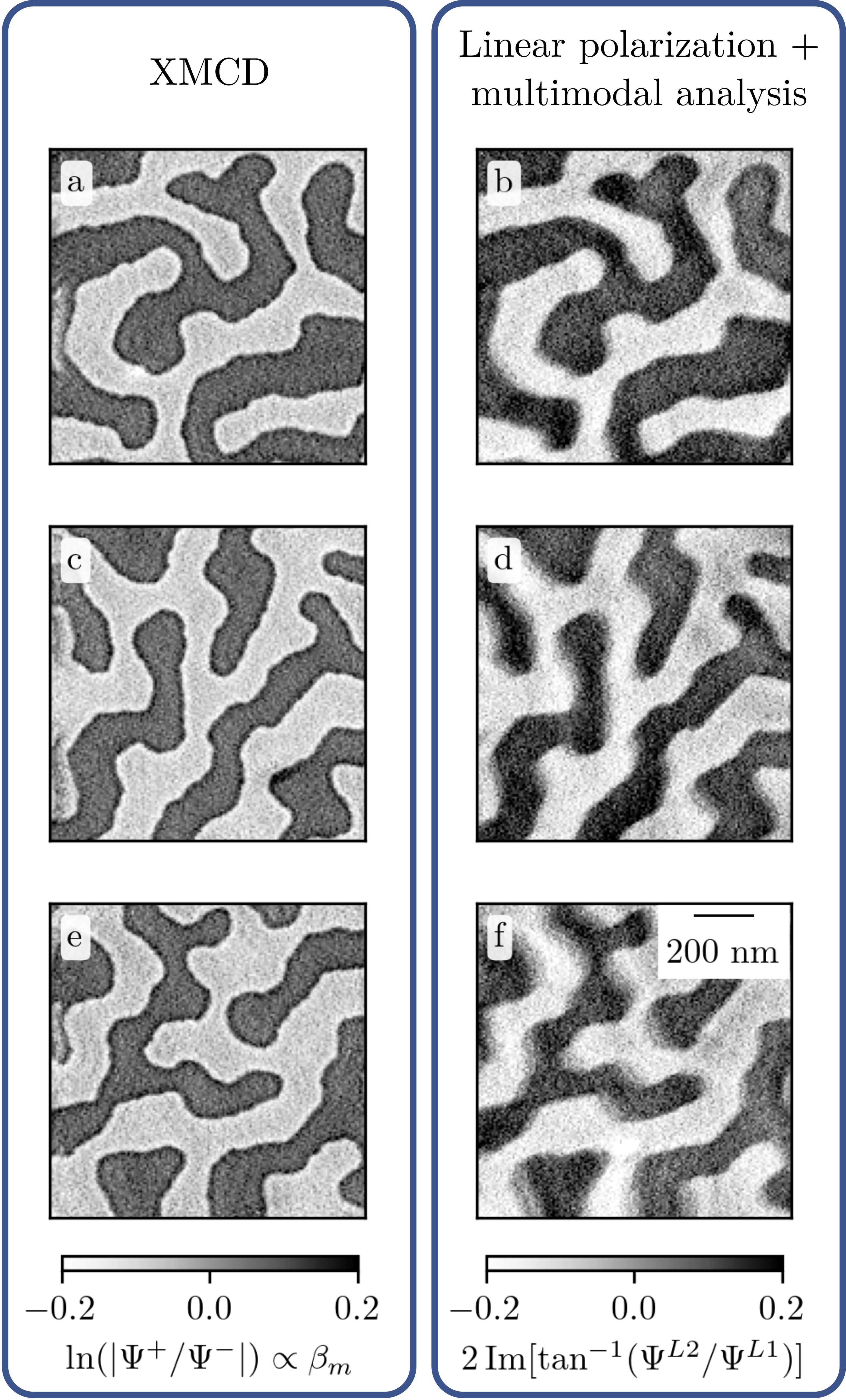}
	\caption{Comparison between the magnetic contrast imaged by XMCD-ptychography (left) and linear polarization with multimodal analysis (right). Worm-like domains are successfully recovered with both methods in a Co/Pt thin film. The field of view is $1\times1$\,$\upmu$m$^2$.}
	\label{fig:cte}
\end{figure}

To demonstrate the method, we started measuring different areas of $1\times1\,\upmu$m$^2$ in the Co/Pt film where the electronic density is uniform.
First, we performed an XMCD-ptychography using both left and right circular polarization to use for comparison and verification. For each polarization, we reconstructed the corresponding object using the ptychography algorithm in PyNX and considering one mode for the object and the probe each.
% By taking the difference between the two polarizations, we were able to probe the out-of-plane magnetization in the thin film. 
We calculate the amplitude contrast following Eq.~\ref{eq:amp} to probe the out-of-plane magnetization in the thin film. 
This can be observed on the left side of Fig.~\ref{fig:cte}. There, worm-like domains of approximately $110$\,nm wide can be observed for three different areas of the film.

Then, we measured a single linear-polarization ptychography in the same areas. In this case, reconstructing the object with only one mode results in a single and uniform object which represents %the topography
% \gb{a chemical contrast (far from absorption edges, it would be the electronic density)}
the electronic density but that is insensitive to the magnetic signal. However, if we enable two modes for the reconstruction of the object, 
% following Eq.~\ref{eq:I2modes},
after orthogonalization of the modes~\cite{thibault2013reconstructing},
and removal of the possible phase ramps and offsets, following Eq.~\ref{eq:imag}, we obtain the magnetic contrast shown 
% we obtain one mode with the uniform electronic structure
% % \gb{chemical} structure
% and another one displaying the magnetization. 
on the right panel of Fig.~\ref{fig:cte}.
% \textbf{RECALCULATE ALL THE NRMSE VALUES FOR THE NEW IMAGES}
It can be seen that the magnetic structure recovered by the multimodal analysis
% is qualitatively equivalent to
matches
the one measured by XMCD-ptychography, that is we can recognize the same shapes and distribution of the domains.
% The spatial resolution estimated in $34$\,nm was calculated from the sharpness of the domain walls and remains the same for both methods.
% \textbf{Compare quantitatively the 2 images.}
% A quantitative comparison was also done by calculating the normalized reduced mean squared error 
% (NRMSE~\footnote{The normalized reduced mean squared error is defined as follows, \begin{equation*} \mbox{NRMSE} = \frac{1}{X_{\mbox{max}} - X_{\mbox{min}}} \left[ \frac{1}{N} \sum_{i=1}^N (X_i - X_{0,i})^2 \right]^{\frac{1}{2}}, \end{equation*} where $X$ and $X_0$ are the reconstructed and the original 3D configuration, respectively, and $N$ is the size of the structure.})
% between the XMCD and the single-polarization-resolved images. This value is less than $12$\% for each of the three images displayed on Fig.~\ref{fig:cte}. 
% We found that the error is mainly concentrated in the flat region of the domains.
% , $40$\% magnetic part and $10$\% the charge for non-uniform.

% \textbf{Resolution?? Why was one bigger than the other? check.} Because to run 2 objects in Wyvern the size must be cut (not enough RAM to process all).
% XMCD: 7nm/px (34nm resolution), Modes: 14.7nm/px (71nm resolution). Resolution calculated with sharpest feature: DWs.

% \textbf{Signal ratio between magnetic and charge part:}

% 1st mode signal: 29740. In reconstruction: 0.038

% 2nd mode signal: 2.57. In reconstruction: 0.0035

% mag/charge signal = 1e-4

% \section{Reconstruction case with non-uniform charge structure}

\begin{figure}
    \includegraphics[width=\columnwidth]{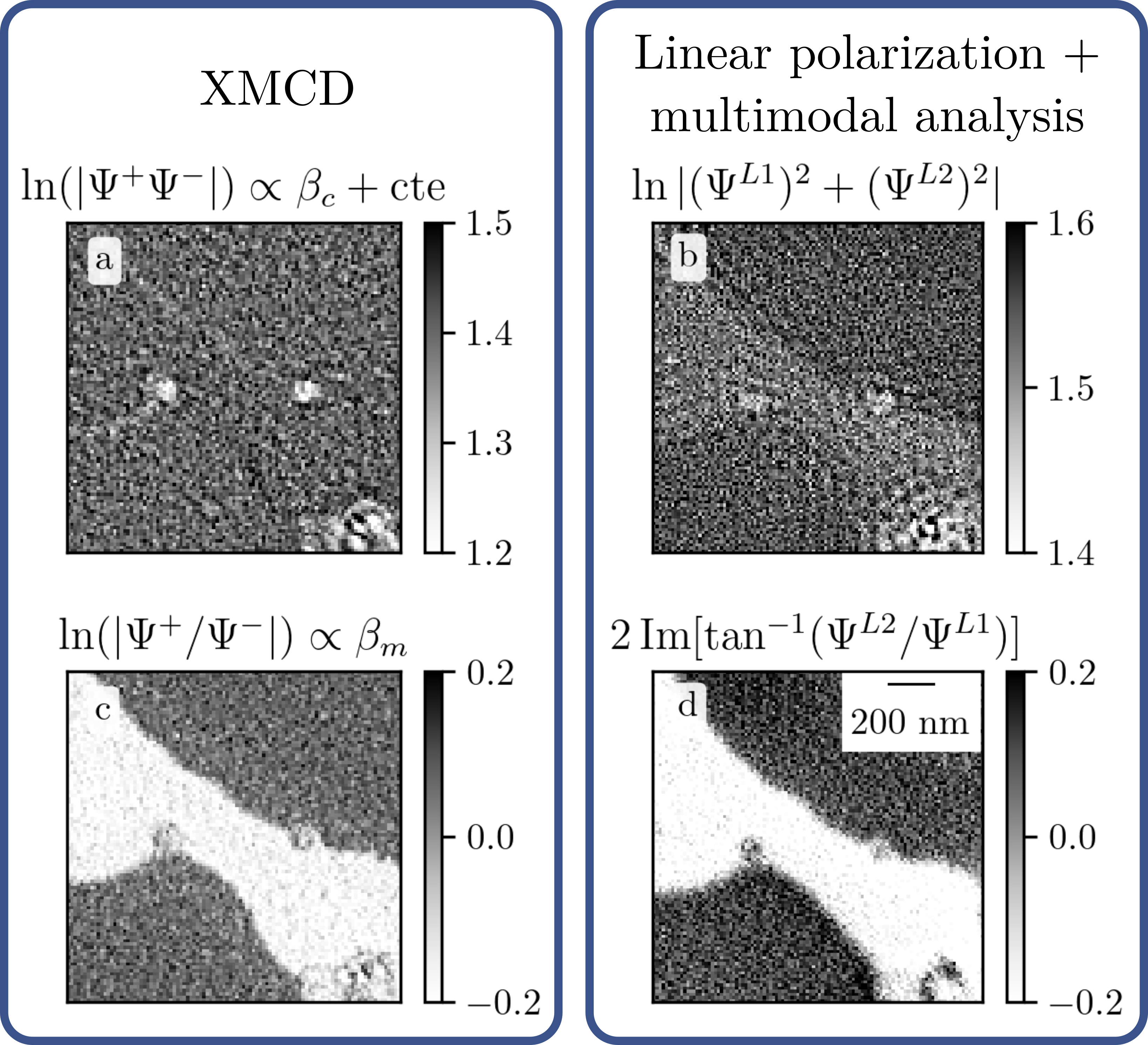}
	\caption{Comparison between XMCD-ptychography (left) and linear polarization with multimodal analysis (right). (a) and (b) correspond to the electronic contrast. There three blobs of carbon can be resolved on the Fe/Gd surface with both methods. The magnetic contrast, shown in (c) and (d), is also captured with both methods. The field of view is $700\times700$\,nm$^2$.}
	\label{fig:noncte}
\end{figure}

Moreover, when the electronic structure 
% \gb{chemical structure} 
is not uniform, we found that disentangling it from the the magnetic signal using the proposed single polarization approach is also possible.
% still possible but more challenging. 
In Fig.~\ref{fig:noncte}, we show the results for
% the first and second modes reconstructed from 
the single linear polarization, %resolved
and we compare it with those from XMCD-ptychography.
% The sum of the objects corresponding to right and left circular polarizations provide contrast for the 
% % \gb{chemical} structure
% electronic structure.
Fig.~\ref{fig:noncte}(a) and (b) show the contrast for the electronic structure.
Two carbon blobs of $52$\,nm diameter are located in the central area and a larger one of $143$\,nm in the lower-right corner of the image. These three features can also be seen in the image obtained from the multimodal analysis of the single linear polarization measurement.
% on the first mode of the single-polarization measurement. 
% Quantitatively, the NRMSE in this case is $10$\%.
On the other hand, 
% the difference between the objects corresponding to right and left circular polarizations, that, as stated above, 
% provides contrast for 
the magnetic contrast displayed in Fig.~\ref{fig:noncte}(c) and (d), shows two opposite domains that are larger than the worm-like domains presented in the previous section. In this case, the single polarization approach has successfully recovered the domain structure.
% walls, although it struggles to display the extension of the domains fully, which corresponds to the lower frequencies. 
% Then, the NRMSE in this case is $40$\%.

% \textbf{Signal ratio between magnetic and charge part:}

% 1st mode signal: 6232.34. In reconstruction: 0.027

% 2nd mode signal: 0.51. In reconstruction: 0.004

% mag/charge signal = 1e-4... it's the same signal ratio as in the uniform case.

\section{Discussion and conclusions}
\label{sec:conclusions}

%%%%%%%%%% What we did
In this work, we demonstrated that, through a multimodal analysis of the reconstructed object, ptychography is able to separate scattered waves with orthogonal polarizations.
Specifically, we applied a multimodal analysis to a single ptychographic measurement performed with linear polarization and retrieved two orthogonal object modes.
The two modes recovered from our experimental data 
reproduce the behavior of the charge scattering on the one hand and the magnetic scattering on the other hand, without further data treatment nor constrains from prior knowledge of the sample.

%%%%%%%%%% Ambiguity, unitary transformation and challenges
In principle, due to the ambiguity in ptychography using multiple object modes, an unitary transformation~\cite{li2016breaking} should be possible to use to optimize the contrast and the separation of the electronic and magnetic signals. 
% Moreover, it turns out that the algorithm naturally chooses the polarization basis that maximises the main mode, which corresponds to a decomposition between charge and magnetic scattering since the former is usually much larger than the latter. 
% This should not necessarily be the case, because any unitary transformation applied to the two modes would fit the experimental measurements equally well. This ambiguity of the multimodal analysis is well known and the choice of a particular pair of solutions relies only on the input of additional information~\cite{li2016breaking}. 
In the case presented here, the ambiguity corresponds to the arbitrary choice of two orthogonal linear polarizations onto which exit waves are projected: the ptychographic algorithm does not know what is the polarization of the incident wave, and that charge scattering should conserve this polarization.
For instance, if the incident polarization is linear horizontal, the set of linear polarizations depicted as $\hat\epsilon_1$ and $\hat\epsilon_2$ in FIG.~\ref{fig:ptycho} is an equivalently good basis from the algorithm point of view. And so would be circular polarizations of opposite helicities, as well as any orthogonal set of elliptical polarizations. 
% A possible explanation as to why the results come out on the relevant basis is that 
However, since 
the charge scattering intensity is larger than the magnetic scattering intensity, 
% in the absence of interference between them,
choosing an initial weight for the second object mode smaller than the first one, guides the decomposition into 
% the mode decomposition possibly maximises the main mode, therefore naturally 
selecting the polarization of charge scattering as one of the vector basis. 
Altogether, the capability of ptychography to resolve orthogonal polarizations with a single measurement and without the need of a physical polarization analyzer
is of great interest for magnetic imaging. In small angle scattering geometry, as presented here, circular dichroism (or birefringence) remains the most efficient way to extract magnetic contrast,
% from the charge information, 
assuming that circular polarization is available. Nevertheless, extracting weak magnetic contrast from circular dichroism relies on a good mutual normalization of the two measurements, which can be difficult in particular when using diffracting quarter-wave plates and on accurate mutual positioning,
while the method presented here relies on a single ptychography scan. 
% Moreover, it turns out that the algorithm naturally chooses the polarization basis that maximises the main mode, which corresponds to a decomposition between charge and magnetic scattering since the former is usually much larger than the latter. 
More importantly, this method 
could be of great interest for magnetic imaging in a wide angle scattering geometry, for instance in diffraction measurements. Indeed, there is no choice of incident polarization allowing to easily extract the magnetic part of the resonant scattering factor from the charge part~\cite{hill1996resonant}. Consequently, it could be an efficient way to image antiferromagnetic textures, including antiphase domains.
% \gb{While the case presented here shows that a single linear polarization can replace circular dichroism measured with two circular polarizations, it should in principle be possible to apply the same method to image linear dichroism / birefringence~\cite{Gao}, of structural or magnetic origin. 
Beyond the cases of circular and linear dichroism and birefringence, where the transmission properties of the sample can be described with only two coefficients, there could be a great interest in combining this method with vectorial ptychography~\cite{Ferrand,Song2020} for the study of anisotropic media in the most general case: four coefficients are then needed to describe the transmission object, and four measurements with a polarization analyzer are needed; the polarization analyzer could be omitted and only two measurements with different incident polarization would be required.
%Furthermore, 
Finally, this method could be applied to other polarized coherent probes than X-rays, and could thus find an application in optical microscopy and specifically in Kerr microscopy.

\section*{Acknowledgments}
% Mention the beamline on which you obtained data, as well as the corresponding proposal number(s):
We acknowledge SOLEIL for providing synchrotron beamtime under project number 20210615.
We acknowledge the Agence Nationale de la Recherche for funding under project number ANR-19-CE42-0013-05 and the CNRS for the grant Emergence@INC2020.
Computations were performed on the HPC system Raven at the Max Planck Computing and Data Facility.

\bibliography{single-pol.bib}
\end{document}